# On spurious regressions with trending variables

MINGHUA WU[†], PIN YOU[†] AND NAN ZOU[‡]

[†]Department of Econometrics and Statistics, School of Economics, Nankai University, Tianjin 300071, China.
E-mail: wminghua_2007@126.com，nkyoupin@126.com

[‡] Department of Mathematics, University of California-San Diego, La Jolla, 92092, United States.
E-mail: nzou@ucsd.edu

**Abstract**

This paper examines three types of spurious regressions where both the dependent and independent variables contain deterministic trends, stochastic trends, or breaking trends. We show that the problem of spurious regression disappears if the trend functions are included as additional regressors. In the presence of autocorrelation, we show that using a Feasible General Least Square (FGLS) estimator can help alleviate or eliminate the problem. Our theoretical results are clearly reflected in finite samples. As an illustration, we apply our methods to revisit the seminal study of Yule (1926).

(This manuscript contains about 9,800 words, including nine tables and one figure.)

## I. Introduction

Spurious regression has attracted much attention in time series econometrics ever since the first simulation studied by Granger and Newbold (1974). A decade later, Phillips (1986) developed the first rigorous asymptotic theory to explain the problem of spurious regression between two independent I(1) processes. Thereafter, spurious regression has been well documented and extended to various scenarios. For example, Marmol (1995) and Tsay and Chung (2000) study the spurious regressions between two independent I(d) processes or long memory processes; Entorf (1997) examines spurious regressions between random walks with drifts. The problem of spurious regression may also occur between independent stationary series, as pointed out by Granger et al. (2001). Kim et al. (2004) consider two statistically independent and highly autocorrelated trend stationary (TS) series and show that usual OLS inference may lead to spurious statistical significance (see also Hasseler, 2000). As an extension, Noriega and Ventosa-Santaulària (2005) and Noriega and Ventosa-Santaulària (2006) show that the phenomenon of spurious regression is also present in TS-I(1) and I(1)-TS regressions. Furthermore, Noriega and Ventosa-Santaulària (2007) show that the spurious regression problem can appear between two time series which exhibit very general nonstationary behaviors. Ventosa-Santaulària (2009) provides an excellent review of these works, and warns that whenever the data-generating processes (DGPs) include a trending mechanism, such as deterministic trends, stochastic trends, or breaking trends, the risk of having a spurious regression is very high.

Spurious regression has been studied extensively as it invalidates the standard statistical inference. Indeed, spurious regression can make two independent random walks look related, but there is actually no relation between them. The statistical relationship may be caused by some lurking variables. Once those lurking variables are taken into consideration, the spurious relationship disappears. Particularly, two deterministic linear trending processes are statistically correlated, since both of them are correlated with a linear trend, as Granger (2012) points out. It has been argued by García-Belmonte and Ventosa-Santaulària (2011) that the spurious regression occurs due to the omission of a lurking variable, so the problem may be solved by introducing a lurking variable as a proxy variable or a trend function in the regression specification. However, they only consider a very simple framework that rules out autocorrelation and structural breaks. Noriega and Ventosa-Santaulària (2006) consider trend stationary time series with breaking trends, and show that including a deterministic trend as an additional regressor does not prevent spurious regression. Our Monte Carlo results also show that simply including a deterministic trend fails to remove the spurious findings of statistical significance, especially when the processes are highly autocorrelated. For this reason, McCallum (2010) advocates using classical autocorrelation corrected procedures such as Cochrane-Orcutt procedure and Feasible Generalized Least Squares (FGLS) to deal with spurious regression. Further, Wu (2013) demonstrates that FGLS can help solve the problem of spurious regression between two stationary autocorrelated processes or two integrated processes (whether or not there is co-integration between them). Martínez-Rivera and Ventosa-Santaulària (2012) show the effectiveness of such a method depends crucially on the DGPs. They argue that the short range autocorrelation should not be considered as the sole source of the spurious regression, other sources, such as deterministic trends, structural breaks, should also be taken into account. To the best of our knowledge, when the problem of a spurious regression is suspected between two trending variables, a general and unified framework to discern the problem is currently lacking.

There is a very general consensus as to the presence of a trending mechanism in the levels of most macroeconomic series. This paper focuses on spurious regression that can be attributed to trending mechanisms. We show that spurious regression can be traced to three sources: the presence of a linear trend, the presence of high autocorrelation, and the presence of breaking trends. In the first case, the spurious regression problem arises from the omission of trend functions in the regression model. When the dependent and the explanatory variables share a common trend, these two variables are spuriously correlated. Spurious regression can be avoided by adding trend functions as explanatory variables. In the second case, the problem arises because we overlook the short range autocorrelation. We can use FGLS to remove the autocorrelation to a great extent. In the third case, the problem arises because we ignore structural breaks. When structural breaks occur, they should be added as explanatory variables to the regression.

This paper studies six cases when the dependent and explanatory variables contain deterministic trends, stochastic trends, or breaking trends, and shows that spurious regression is present in all cases. By not only controlling for the lurking variables such as trends and structural breaks in the specification, but also using FGLS in the presence of autocorrelation, we develop a unifying approach to identify the problem of spurious regression. It is shown that the $t$-statistic converges weakly to the standard normal distribution or a functional of the Wiener process. Our method can thus alleviate or eliminate the spurious regression problem. Further, simulation experiments reveal that the phenomenon of spurious regression has been eliminated, confirming the effectiveness of our method. Finally, we apply our method to analyze the relationship between the two time series in Yule (1926).

The rest of the article is organized as follows. Section 2 describes the DGPs and regression specifications. Section 3 describes the spurious regressions and our proposed method to deal with them. We establish the asymptotic distributions of the test statistics and provide simulation evidence for each case. Section 4 contains the empirical study. Section 5 provides some concluding comments. Appendix A provides the proofs of the four theorems given in the main text, and Appendix B gives some supplementary results.

## II. DGPs and specifications

Spurious regression is well known to be present under different forms of non-stationarity in the DGP. Specifically, when fitting two non-stationary and independent time series $y_t$ and $x_t$ in a regression model $y_t = \alpha + \gamma x_t + u_t$, the OLS estimator of $\gamma$ does not converge to its true value of zero, and the *t*-statistic for testing the hypothesis $H_0: \gamma = 0$ diverges, thus indicating the presence of a spurious asymptotic relationship between $y_t$ and $x_t$.

In general, the nature of the trending mechanism in $y_t$ and $x_t$ is unknown *a priori* in a regression model. For this reason, we consider three possible combinations of non-stationarity. In Case 1, denoted by TS-TS, we regress a trend-stationary series on another trend-stationary series. In Case 2, denoted by TS-I(1), we regress a trend-stationary series on an integrated process. In Case 3, denoted by I(1)-TS, we regress an integrated process on a trend-stationary series, The innovations in each case are allowed to be autocorrelated. Drifts are also allowed. Each case is divided into two sub-cases depending on whether structural breaks are allowed or not --- Case A does not have any structural break, and Case B has some structural breaks.

The assumptions in TABLE 1 summarize the DGPs for both the dependent and the explanatory variables. In TABLE 1, $\varepsilon_{yt}$ and $\varepsilon_{xt}$ are independent innovations, $\varepsilon_{yt} \sim$ iid $(0, \sigma_y^2)$, $\varepsilon_{xt} \sim$ iid $(0, \sigma_x^2)$, $\mu$, $\beta$, and $\phi$ are constants. *DU* and *DT* are dummy variables indicating trend and slope breaks, for example, $DU_{iyt} = 1(t > T_{biy})$, $i = 1,2,\ldots,N_y$, $DT_{iyt} = (t - T_{biy})1(t > T_{biy})$, $i = 1,2,\ldots,M_y$, where *N* and *M* are positive integers, $1(\cdot)$ is the indicator function, and $T_{biy}$ is the location of the structural break.

The specifications for regression models are given in TABLE 2. The five specifications are used as a vehicle to study the phenomenon of spurious regressions and solutions. Regression 1 is the simplest case. From Regression 2 to Regression 5, a linear time trend is included as an additional regressor which happens to be the trending mechanism. Regression 3 applies FGLS. Regression 4 includes structural breaks in the equation; Regression 5 applies FGLS and includes structural breaks. FGLS is to proceed in two stages: (1) we estimate the model by OLS or another consistent (but inefficient) estimator, and build a consistent estimator of the errors covariance matrix using the residuals; (2) we implement generalized least squares (GLS) to estimate the unknown parameters using the consistent estimator of the errors covariance matrix.

TABLE 1
*Assumptions: the DGPs for $y_t$ and $x_t$*

| Case | Series | $y_t$ and $x_t$ |
|---|---|---|
| 1A | TS | $y_t = \mu_y + \beta_y t + u_{yt}$; $u_{yt} = \phi_y u_{yt-1} + \varepsilon_{yt}$, $|\phi_y|<1$ |
|  | TS | $x_t = \mu_x + \beta_x t + u_{xt}$; $u_{xt} = \phi_x u_{xt-1} + \varepsilon_{xt}$, $|\phi_x|<1$ |
| 1B | TS+br | $y_t = \mu_y + \sum_{i=1}^{N_y}\mu_{iy}DU_{iyt} + \beta_y t + \sum_{i=1}^{M_y}\beta_{iy}DT_{iyt} + u_{yt}$; $u_{yt} = \phi_y u_{yt-1} + \varepsilon_{yt}$, $|\phi_y|<1$ |
|  | TS+br | $x_t = \mu_x + \sum_{i=1}^{N_x}\mu_{ix}DU_{ixt} + \beta_x t + \sum_{i=1}^{M_x}\beta_{ix}DT_{ixt} + u_{xt}$; $u_{xt} = \phi_x u_{xt-1} + \varepsilon_{xt}$, $|\phi_x|<1$ |
| 2A | TS | $y_t = \mu_y + \beta_y t + u_{yt}$; $u_{yt} = \phi_y u_{yt-1} + \varepsilon_{yt}$, $|\phi_y|<1$ |
|  | I(1) | $x_t = \beta_x + x_{t-1} + \varepsilon_{xt}$ |
| 2B | TS+br | $y_t = \mu_y + \sum_{i=1}^{N_y}\mu_{iy}DU_{iyt} + \beta_y t + \sum_{i=1}^{M_y}\beta_{iy}DT_{iyt} + u_{yt}$; $u_{yt} = \phi_y u_{yt-1} + \varepsilon_{yt}$, $|\phi_y|<1$ |
|  | I(1)+br | $x_t = \beta_x + \sum_{i=1}^{N_x}\beta_x DU_{ixt} + x_{t-1} + \varepsilon_{xt}$ |
| 3A | I(1) | $y_t = \beta_y + y_{t-1} + \varepsilon_{yt}$ |
|  | TS | $x_t = \mu_x + \beta_x t + u_{xt}$; $u_{xt} = \phi_x u_{xt-1} + \varepsilon_{xt}$, $|\phi_x|<1$ |
| 3B | I(1)+br | $y_t = \beta_y + \sum_{i=1}^{N_y}\beta_{iy}DU_{iyt} + y_{t-1} + \varepsilon_{yt}$ |
| TS+br | $x_t = \mu_x + \sum_{i=1}^{N_x} \mu_{ix} DU_{ixt} + \beta_x t + \sum_{i=1}^{M_x} \beta_{ix} DT_{ixt} + u_{xt};$ $u_{xt} = \phi_x u_{xt-1} + \varepsilon_{xt}, |\phi_x| < 1$ | |

Notes: *TS* and *br* denote "trend stationary" and "structural break" respectively.

TABLE 2
Specifications for regression models

| No. of Regression | Specification | Description |
|---|---|---|
| 1 | $y_t = \hat{\alpha} + \hat{\gamma} x_t + \hat{e}_t$ | The simplest OLS regression |
| 2 | $y_t = \hat{\alpha} + \hat{\beta} t + \hat{\gamma} x_t + \hat{e}_t$ | OLS regression with time as a regressor |
| 3 | $y_t = \hat{\alpha} + \hat{\beta} t + \hat{\gamma} x_t + u_t, \ u_t = \hat{\rho} u_{t-1} + \hat{e}_t$ | FGLS regression with time as a regressor |
| 4 | $y_t = \hat{\alpha} + \hat{\beta} t + \hat{\beta}_1 DT_{yt} + \hat{\gamma} x_t + \hat{e}_t$ | OLS regression with time and structural breaks as regressors |
| 5 | $y_t = \hat{\alpha} + \hat{\beta} t + \hat{\beta}_1 DT_{yt} + \hat{\gamma} x_t + u_t, \ u_t = \hat{\rho} u_{t-1} + \hat{e}_t$ | FGLS regression with time and structural breaks as regressors |

## III. Solution to spurious regression problem: asymptotic and simulation evidence

This section presents the asymptotic behavior of the estimator $\hat{\gamma}$, and its *t*-statistic $t_{\hat{\gamma}}$. The main interest lies in the *t*-statistic $t_{\hat{\gamma}}$. If the regression yields *t*-statistic with large absolute value, despite the independence between the variables, the phenomenon of spurious regression is present.

**The case of regression between two trend stationary variables (TS-TS)**

*Case 1A (TS-TS regression without structural breaks)*

Given two independent stationary processes $y_t$ and $x_t$ around trends, which are generated from the DGPs as in Case 1A in TABLE 1, Kim et al. (2004) present analytical and empirical evidences that the phenomenon of spurious regression occurs in Regression 1. Although $y_t$ and $x_t$ are totally unrelated trending time series, there is a highly significant relationship between them, because they are driven by a common linear trend.

As discussed above, the spurious relationship may be caused by some lurking variables.

García-Belmonte and Ventosa-Santaulària (2011) argue that the existence of a deterministic trend in the variables seems to underlie the phenomenon of spurious regression in many cases and show that it can be eliminated when the trending mechanism is included as a regressor. More specifically, by adding a time trend to Regression 1, they get Regression 2 and shows that, when $\phi_y = \phi_x = 0$, the limiting distribution of $t_{\hat{\gamma}}$ is standard normal. Also, they provide some Monte Carlo evidence that the spurious relationship disappears regardless of the sample size. However, Regression 2 cannot deal with spurious regression in finite samples when $y_t$ has first-order autocorrelation, as the stronger the autocorrelation, the severer the spurious regression problem. To improve the approach given by García-Belmonte and Ventosa-Santaulària (2011), we take into account not only the lurking trend variable but also the autocorrelation. Specifically, by using FGLS in Regression 3, we can avoid the spurious regression problem. The following proposition presents the asymptotics in this case.

*Proposition 1*. Suppose $y_t$ and $x_t$, $t=1, 2,..., T$ are independent trend stationary series generated by Case 1A in TABLE 1, and $\phi_y \neq 0$. Running Regression 3 with FGLS, we get

$$\hat{\gamma} \xrightarrow{d} N(0, \sigma_{\hat{\gamma}}^2) , \quad t_{\hat{\gamma}} \xrightarrow{d} N(0,1)$$

where $\sigma_{\hat{\gamma}}^2$ depends on $\phi_y$, $\sigma_y^2$ and $x_t$.

*Proof.* See Appendix A.

Proposition 1 shows when applying FGLS to the regression involving two trend stationary variables, the estimator $\hat{\gamma}$ converges to zero and is asymptotically normal, while its *t*-statistic $t_{\hat{\gamma}}$ is asymptotically standard normal. This indicates that Regression 3, which uses FGLS with linear trend as an additional regressor, can eliminate spurious regression between two trend stationary variables.

Now we investigate the finite sample behavior of the *t*-statistic $t_{\hat{\gamma}}$ under Regression 2 and Regression 3 by Monte Carlo simulations. The DGPs of $y_t$ and $x_t$ are the Case 1A in TABLE 1, where the two error terms $\varepsilon_{yt}$ and $\varepsilon_{xt}$ are drawn from $N(0,1)$. Other parameters are chosen to be identical with those in Kim et al. (2004) for comparison purposes. We apply Regression 2 when the dependent variable $y_t$ has no autocorrelation (when $\phi_y = 0$), and use Regression 3 otherwise (when $\phi_y \neq 0$). The number of replications is 10,000. TABLE 3 reports the rejection rates of the null hypothesis $H_0: \gamma = 0$ at the 5% significant level.

TABLE 3
*Proportion of rejections of Case 1A (TS-TS regression without structural breaks)*

| $\beta_y$ | $\beta_x$ | $\phi_y$ | $\phi_x$ | Regression 1 | | | Regression 2 or 3 | | |
|---|---|---|---|---|---|---|---|---|---|
| | | | | T=50 | T=100 | T=10000 | T=50 | T=100 | T=10000 |
| 0 | 0 | 0 | 0 | 0.0575 | 0.0527 | 0.0515 | 0.0508 | 0.0529 | 0.0519 |
| | | 0.3 | 0.3 | 0.0806 | 0.0741 | 0.0712 | 0.0548 | 0.0531 | 0.0485 |
| | | 0.9 | 0.9 | 0.4445* | 0.5041* | 0.5293* | 0.0696 | 0.0598 | 0.0493 |
| | | 0 | 0.9 | 0.0529 | 0.0518 | 0.0504 | 0.0509 | 0.0522 | 0.0505 |
| | | 0.9 | 0 | 0.0550 | 0.0497 | 0.0447 | 0.0453 | 0.0422 | 0.0457 |
| 0 | 0.2 | 0 | 0 | 0.0414 | 0.0492 | 0.0496 | 0.0511 | 0.0502 | 0.0497 |
| | | 0.3 | 0.3 | 0.1300* | 0.1469* | 0.1520* | 0.0538 | 0.0516 | 0.0506 |
| | | 0.9 | 0.9 | 0.5827* | 0.6307* | 0.6558* | 0.0689 | 0.0600 | 0.0494 |
| | | 0 | 0.9 | 0.0466 | 0.0516 | 0.0508 | 0.0465 | 0.0501 | 0.0501 |
| | | 0.9 | 0 | 0.5993* | 0.6321* | 0.6541 * | 0.0392 | 0.0440 | 0.0491 |
| 0.2 | 0 | 0 | 0 | 0.0436 | 0.0490 | 0.0464 | 0.0518 | 0.0518 | 0.0511 |
| | | 0.3 | 0.3 | 0.1261* | 0.1431* | 0.1482* | 0.0579 | 0.0534 | 0.0538 |
| | | 0.9 | 0.9 | 0.5706* | 0.6311* | 0.6554* | 0.0660 | 0.0527 | 0.0529 |
| | | 0 | 0.9 | 0.6020* | 0.6340* | 0.6542* | 0.0474 | 0.0478 | 0.0517 |
| | | 0.9 | 0 | 0.0441 | 0.0503 | 0.0474 | 0.0434 | 0.0469 | 0.0566 |
| 0.2 | 0.2 | 0 | 0 | 1.0000* | 1.0000* | 1.0000* | 0.0524 | 0.0513 | 0.0518 |
| | | 0.3 | 0.3 | 1.0000* | 1.0000* | 1.0000* | 0.0548 | 0.0508 | 0.0505 |
| | | 0.9 | 0.9 | 0.9826* | 1.0000* | 1.0000* | 0.0662 | 0.0570 | 0.0478 |
| | | 0 | 0.9 | 0.9953* | 1.0000* | 1.0000* | 0.0486 | 0.0483 | 0.0508 |
| | | 0.9 | 0 | 0.9963* | 1.0000* | 1.0000* | 0.0405 | 0.0452 | 0.0551 |

Notes: * indicates that the proportion of rejections is greater than 10%, which provides evidence on the presence of a spurious regression.

*Remark 1*. We also concern about the situation using the heteroscedasticity and autocorrelation consistent (HAC) standard error (which is also known as the *Newey-West* standard error). Regression 2 with *Newey-West* standard error (hereinafter Regression 2(*NW*)) can alleviate the spurious regression problem in large samples (see Gujarati and Porter, 2009, pp. 447-448). However, if the sample size is not large enough, as in the present context, Regression 2(*NW*) is less efficient than the FGLS method in Regression 3. By Monte Carlo simulation, we compare the rejection rates of the null hypothesis $H_0$: $\gamma = 0$ at the 5% significant level, under Regression 2, Regression 2(*NW*), and Regression 3 respectively. The results are listed in Table B1 in Appendix B. The sample size is 100, and other parameters are matched with those in TABLE 3 . It can be seen that the rejection rates under Regression 2(*NW*) are all higher than those under Regression 3. Particularly, when both series are highly autocorrelated ($\phi_y=\phi_x=0.9$), the rejection rates under Regression 3 are close to the 5% nominal significant level (6.69% and 6.62% for $\beta_y=0$, $\beta_x=0.2$ and $\beta_y=0.2$, $\beta_x=0.2$, respectively), while the rejection rates under Regression 2(*NW*) and Regression 2 are much higher than the level 5% (both of the value are over 30% and 40%, respectively). So we employ FGLS method rather than HAC, to correct the OLS standard error when it is autocorrelated. It is worth noting that there are some new developments about HAC standard error (Jin et al., 2006; Müller, 2007; Sun, 2004, 2014a, 2014b). However, as Sun (2014a) puts forward, a good method may require some prior knowledge about the data generating process.

*Case1B (TS-TS regression with structural breaks)*

Noriega and Ventosa-Santaulària (2006) extend the work of Kim et al. (2004) in order to study the impact of structural breaks on the spurious regressions between two trend stationary series. They prove that the phenomenon of spurious regression is present when $y_t$ and $x_t$ are trend stationary processes with single or multiple structural breaks.

To solve the problem, note that spurious regression is also originated from lurking variables——not only the common trend, but also the structural breaks. Therefore, when structural breaks occur, the spurious regression between trend stationary series can be removed by including the trend variable and the structural breaks as regressors.

To confirm this, we run Monte Carlo experiments by considering the simple case with at most one structural break. For simplicity, we assume that the number, the types, and the locations of the structural breaks are known. If they are unknown, we should estimate them first. The same comment applies to other settings with structural breaks.

In the case that $y_t$ has no structural break, we apply Regression 2 or Regression 3 depending on whether $\phi_y$ is 0 or not. In the case that $y_t$ involves a structural break, we choose Regression 4 or Regression 5 depending on whether $\phi_y$ is 0 or not. The trend parameters are chosen to be $\beta_y=\beta_x=0.2$. The choices of other parameters are given in TABLE 4. The break in $y_t$ is located at $T/2$, and the break in $x_t$ is at $T/5$, where $T$ is selected to be 50, 100, and 10,000. The number of replications is 10,000. TABLE 4 reports the rejection rates of the null hypothesis H$_0$: $\gamma = 0$ at the 5% significant level.

TABLE 4
*Proportion of rejections of Case 1B (TS-TS regression with structural breaks)*

| $\beta_{1y}$ | $\beta_{1x}$ | $\phi_y$ | $\phi_x$ | Regression 1 | | | Regression 2-5 | | |
|---|---|---|---|---|---|---|---|---|---|
| | | | | $T=50$ | $T=100$ | $T=10000$ | $T=50$ | $T=100$ | $T=10000$ |
| 0 | 0.2 | 0 | 0 | 1.0000 | 1.0000 | 1.0000 | 0.0505 | 0.0478 | 0.0469 |
| | | 0.3 | 0.3 | 1.0000 | 1.0000 | 1.0000 | 0.0618 | 0.0512 | 0.0457 |
| | | 0.9 | 0.9 | 0.9893 | 1.0000 | 1.0000 | 0.0754 | 0.0548 | 0.0495 |
| | | 0 | 0.9 | 0.9930 | 1.0000 | 1.0000 | 0.0484 | 0.0516 | 0.0526 |
| | | 0.9 | 0 | 1.0000 | 1.0000 | 1.0000 | 0.0419 | 0.0417 | 0.0502 |
| 0.2 | 0 | 0 | 0 | 1.0000 | 1.0000 | 1.0000 | 0.0519 | 0.0438 | 0.0517 |
| | | 0.3 | 0.3 | 1.0000 | 1.0000 | 1.0000 | 0.0602 | 0.0561 | 0.0454 |
| | | 0.9 | 0.9 | 0.9953 | 1.0000 | 1.0000 | 0.0706 | 0.0554 | 0.0532 |
| | | 0 | 0.9 | 1.0000 | 1.0000 | 1.0000 | 0.0511 | 0.0532 | 0.0447 |
| | | 0.9 | 0 | 0.9963 | 1.0000 | 1.0000 | 0.0413 | 0.0407 | 0.0500 |
| 0.2 | 0.2 | 0 | 0 | 1.0000 | 1.0000 | 1.0000 | 0.0458 | 0.0454 | 0.0467 |
| | | 0.3 | 0.3 | 1.0000 | 1.0000 | 1.0000 | 0.0602 | 0.0560 | 0.0499 |
| | | 0.9 | 0.9 | 1.0000 | 1.0000 | 1.0000 | 0.0688 | 0.0560 | 0.0533 |
| | | 0 | 0.9 | 1.0000 | 1.0000 | 1.0000 | 0.0489 | 0.0469 | 0.0509 |
| | | 0.9 | 0 | 1.0000 | 1.0000 | 1.0000 | 0.0408 | 0.0483 | 0.0500 |

As can be seen from TABLE 4, under Regression 1, the finite sample rejection rates are 100%. However, this spurious regression can be removed successfully by using Regressions 2-5. In summary, the problem of spurious regression between two trend stationary variables can be resolved in the following way. We add the trending mechanism as a regressor first. If the error is autocorrelated, we employ the FGLS. In this case, the estimator $\hat{\gamma}$ is asymptotically normal, and its *t*-statistic $t_{\hat{\gamma}}$ is asymptotically standard normal. When $y_t$ involves structural breaks, the trending mechanism and the structural breaks should be added as regressors, and OLS or FGLS should be used depending on whether the error is uncorrelated or not.

**Spurious regression of a trend stationary process on an integrated process (TS-I (1))**

*Case 2A (TS-I (1) regression without structural breaks)*

Consider two independent series $y_t$ and $x_t$ generated by Case 2A in TABLE 1, where $y_t$ is a trend stationary process and $x_t$ is an integrated process. Noriega and Ventosa-Santaulària (2007) show that the *t*-statistic for testing a linear relationship between independent time series (i.e., $t_{\hat{\gamma}}$) diverges, so the phenomenon of spurious regression is present.

In light of this, we now present a solution to the problem of spurious regression when a trend stationary process is regressed on an integrated process. Note that although the integrated process may not have a trend in itself, it does have spurious correlation with time due to its high persistency (see Durlauf and Phillips, 1988). Therefore, both $y_t$ and $x_t$ are correlated with time, just as in Case 1. Hence, under Regression 1 spurious regression occurs. This problem can be tackled by adding time as a regressor. Proposition 2 below shows that Regression 2 has solved the spurious regression problem when the error is not autocorrelated.

*Proposition 2.* Suppose $y_t$ and $x_t$ are independent series generated by Case2A in TABLE 1, and $\phi_y=0$. Running Regression 2, we get

$$T\hat{\gamma} \Rightarrow \frac{\frac{1}{12}\sigma_v[\int_0^1 W(t)dV(t) - V(1)\int_0^1 W(t)dt] - \sigma_v[\frac{1}{2}V(1) - \int_0^1 V(t)dt][\int_0^1 tW(t)dt - \frac{1}{2}\int_0^1 W(t)dt]}{\frac{1}{12}\sigma_w\{\int_0^1 W(t)^2 dt - [\int_0^1 W(t)dt]^2\} - \sigma_w\{\int_0^1 tW(t)dt - \frac{1}{2}\int_0^1 W(t)dt\}^2}$$

$$t_{\hat{\gamma}} \Rightarrow \frac{\frac{1}{12}[\int_0^1 W(t)dV(t) - V(1)\int_0^1 W(t)dt] - [\frac{1}{2}V(1) - \int_0^1 V(t)dt][\int_0^1 tW(t)dt - \frac{1}{2}\int_0^1 W(t)dt]}{\sqrt{\frac{1}{12^2}\{\int_0^1 W(t)^2 dt - [\int_0^1 W(t)dt]^2\} - \frac{1}{12}\{\int_0^1 tW(t)dt - \frac{1}{2}\int_0^1 W(t)dt\}^2}}$$

where $\Rightarrow$ denotes weak convergence, and $W(t)$ and $V(t)$ denote two independent Wiener processes on $C[0,1]$.

*Proof.* See Appendix A.

Proposition 2 indicates that the *t*-statistics $t_{\hat{\gamma}}$ converges weakly to a limiting distribution free of unknown parameters. Although this limiting distribution is a functional of Wiener processes and not precisely the standard normal distribution, it is very close to the standard normal distribution (see Noriega and Ventosa-Santaulària, 2005), and so its quantiles can be well approximated by the standard normal quantiles. With this approximation, we can implement the *t*-test and thus solve the spurious regression problem.

Further, if $y_t$ contains autocorrelation, that is, the parameter $\phi_y$ in Case 2A in TABLE 1 does not equal to zero, then besides adding the trending mechanism as a regressor, we also need to implement FGLS to deal with the autocorrelation. Hence we use Regression 3, an approach justified by Proposition 3.

*Proposition 3.* Suppose independent series $y_t$ and $x_t$ are generated by Case2A in TABLE 1, and $\phi_y \neq 0$. Running regression 3 with FGLS, we get

$$T\hat{\gamma} \Rightarrow \frac{\sigma_y\{\frac{1}{12}[\int_0^1 W(t)dV(t) - V(1)\int_0^1 W(t)dt] - [\frac{1}{2}V(1) - \int_0^1 V(t)dt][\int_0^1 tW(t)dt - \frac{1}{2}\int_0^1 W(t)dt]\}}{((1-\phi_y)\sigma_w)\{\frac{1}{12}\{\int_0^1 W(t)^2 dt - [\int_0^1 W(t)dt]^2\} - \{\int_0^1 tW(t)dt - \frac{1}{2}\int_0^1 W(t)dt\}^2\}}$$

$$t_{\hat{\gamma}} \Rightarrow \frac{\{\frac{1}{12}[\int_0^1 W(t)dV(t) - V(1)\int_0^1 W(t)dt] - [\frac{1}{2}V(1) - \int_0^1 V(t)dt][\int_0^1 tW(t)dt - \frac{1}{2}\int_0^1 W(t)dt]\}}{\sqrt{\frac{1}{12^2}\{\int_0^1 W(t)^2 dt - [\int_0^1 W(t)dt]^2\} - \frac{1}{12}\{\int_0^1 tW(t)dt - \frac{1}{2}\int_0^1 W(t)dt\}^2}}$$

where $\Rightarrow$ denotes weak convergence, and $W(t)$ and $V(t)$ are independent Wiener processes on $C[0,1]$.

*Proof.* See Appendix A.

Proposition 3 shows that the *t*-statistics $t_{\hat{\gamma}}$ converges weakly to a limiting distribution free of unknown parameters. This limiting distribution is identical with that in Proposition 2, which is very close to the standard normal distribution. Hence, the quantiles of the limiting distribution can be approximated by the standard normal quantiles. With this approximation, the *t*-test can be implemented and thus the spurious regression problem can be solved.

Here we need to note that, when the error is not autocorrelated, running Regression 3 with FGLS, we will get the same limiting distributions for the coefficient estimator and the *t*-statistic. Hence, Regression 3 with FGLS can be applied regardless of whether autocorrelation is present. However, when the error is not autocorrelated, running FGLS introduces an extra estimation error and leads to upward size distortion in finite samples. The choice between OLS and FGLS depends on the autocorrelation structure of the innovations, as Wu (2013) suggested.

TABLE 5
*Proportion of rejections of Case 2A (TS-I (1) regression without structural breaks)*

| $\beta_y$ | $\beta_x$ | $\phi_y$ | Regression 1 | | | Regression 2 or 3 | | |
|---|---|---|---|---|---|---|---|---|
| | | | T=50 | T=100 | T=10000 | T=50 | T=100 | T=10000 |
| 0 | 0 | 0 | 0.0484 | 0.0464 | 0.0514 | 0.0477 | 0.0488 | 0.0525 |
| | | 0.3 | 0.1315* | 0.1442* | 0.1485* | 0.0736 | 0.0626 | 0.0507 |
| | | 0.9 | 0.5317* | 0.5923* | 0.6564* | 0.0782 | 0.0648 | 0.0489 |
| 0 | 0.2 | 0 | 0.0467 | 0.0475 | 0.0486 | 0.0509 | 0.0482 | 0.0525 |
| | | 0.3 | 0.1348* | 0.1427* | 0.1539* | 0.0789 | 0.0677 | 0.0491 |
| | | 0.9 | 0.5533* | 0.6145* | 0.6552* | 0.0820 | 0.0719 | 0.0526 |
| 0.2 | 0 | 0 | 0.8098* | 0.8775* | 0.9867* | 0.0523 | 0.0516 | 0.0472 |
| | | 0.3 | 0.8149* | 0.8810* | 0.9882* | 0.0797 | 0.0669 | 0.0537 |
| | | 0.9 | 0.7824* | 0.8653* | 0.9887* | 0.0828 | 0.0746 | 0.0497 |
| 0.2 | 0.2 | 0 | 0.9186* | 0.9732* | 1.0000* | 0.0518 | 0.0478 | 0.0500 |
| | | 0.3 | 0.9182* | 0.9783* | 1.0000* | 0.0787 | 0.0668 | 0.0485 |
| | | 0.9 | 0.8962* | 0.9718* | 1.0000* | 0.0872 | 0.0739 | 0.0512 |

Notes: * indicates that the proportion of rejections is more than 10%, which shows the phenomenon of spurious regression.

Now we investigate the finite sample behavior of the *t*-statistic $t_{\hat{\gamma}}$ by conducting Monte Carlo simulations. $y_t$ and $x_t$ are generated by Case 2A in TABLE 1, where $\varepsilon_{yt}$ and $\varepsilon_{xt}$ have standard normal distributions, $\mu_y$ is chosen to be 0.8 (The simulation results have nothing to do with the selection of $\mu_y$, and it will be the same hereafter), and the choices of other parameters are given in TABLE 5. Regression 2 or Regression 3 is applied depending on whether the error is uncorrelated or not. The number of replications is 10,000. TABLE 5 reports the rejection rates of the null hypothesis $H_0$: $\gamma = 0$ at the 5% significant level.

Under Regression 1, if $y_t$ has a deterministic trend ($\beta_y \neq 0$), the probability of spurious regression is very high, regardless of whether $x_t$ contains a drift or not. Moreover, if $y_t$ has autocorrelation, then as $\phi$ approaches 1, the rejection rate becomes larger. For example, the finite sample rejection rates for the case of $\beta_y=\beta_x=0.2$ and $T=100$, are 59% and 61% for $\phi_y= 0.3$ and 0.9, respectively.

On the other hand, the remaining experiments by Regression 2 and Regression 3 reveal that the rejection rates are very close to the nominal 5% level, which suggests that the problem of spurious regression has been eliminated successfully.

*Case 2B (TS-I(1) regression with structural breaks ).*

Noriega and Ventosa-Santaulària (2007) take structural breaks into accounts when using regression 1 to regress a trend stationary process on an integrated process. Consider two independent time series $y_t$ and $x_t$ generated by Case 2B in TABLE 1, where $y_t$ is a trend stationary process, $x_t$ is an integrated process, and both may involve breaks in intercepts and/or in slopes. It turns out that under Regression 1, the *t*-statistic diverges with rate $O(T^{1/2})$, so the phenomenon of spurious regression is present.

We use the same idea as above to deal with the spurious regression problem, by including the trending mechanism and structural breaks as regressors, as well as using FGLS when the error is autocorrelated.

We present the Monte Carlo performance of those *t*-statistics, considering the simple case that each of $y_t$ and $x_t$ contains at most one structural break. If the dependent variable $y_t$ has no structural break, we apply Regression 2 or Regression 3 depending on whether the error is uncorrelated or not. If $y_t$ involves a structural break, we choose Regression 4 or Regression 5 depending on whether the error is uncorrelated or not. The trend parameters in our simulations are chosen to be $\beta_y = \beta_x = 0.2$. The choices of other parameters are given in TABLE 6. The break in $y_t$ is located at $T/2$, and the break in $x_t$ is at $T/5$, where $T$ is selected to be 50, 100, and 10,000. The number of replications is 10,000. TABLE 6 shows the rejection rates of the null hypothesis $H_0: \gamma = 0$ at the 5% significant level.

TABLE 6
*Proportion of rejections of Case 2B (TS-I(1) regression with structural breaks)*

| $B_{1y}$ | $\beta_{1x}$ | $\phi_y$ | Regression 1 | | | Regression 2-5 | | |
|---|---|---|---|---|---|---|---|---|
| | | | T=50 | T=100 | T=10000 | T=50 | T=100 | T=10000 |
| | | 0 | 0.9938 | 0.9998 | 1.0000 | 0.0528 | 0.0497 | 0.0505 |
| 0 | 0.2 | 0.3 | 0.9950 | 0.9999 | 1.0000 | 0.0803 | 0.0637 | 0.0491 |
| | | 0.9 | 0.9870 | 0.9999 | 1.0000 | 0.0869 | 0.0695 | 0.0536 |
| | | 0 | 0.9114 | 0.9742 | 1.0000 | 0.0462 | 0.0514 | 0.0485 |
| 0.2 | 0 | 0.3 | 0.9125 | 0.9761 | 1.0000 | 0.0822 | 0.0704 | 0.0500 |
| | | 0.9 | 0.9125 | 0.9734 | 1.0000 | 0.0923 | 0.0737 | 0.0536 |
| | | 0 | 0.9932 | 0.9999 | 1.0000 | 0.0472 | 0.0495 | 0.0542 |
| 0.2 | 0.2 | 0.3 | 0.9927 | 1.0000 | 1.0000 | 0.0818 | 0.0672 | 0.0488 |
| | | 0.9 | 0.9933 | 1.0000 | 1.0000 | 0.0875 | 0.0707 | 0.0467 |

TABLE 6 confirms that the phenomenon of spurious regression is present under Regression 1. Moreover, it has been eliminated under Regressions 2-5. To achieve even better finite sample results when $y_t$ is serially correlated, one may consider using the feasible generalized median estimator.

In summary, the spurious regression of a trend stationary process on an integrated process can be removed in the following way. We add the trending mechanism as a regressor first; if the error is autocorrelated, we employ the FGLS (as Regression 3). In this case, the *t*-statistic $t_{\hat{\gamma}}$ converges to a functional of Wiener process which is close to be standard normal. When $y_t$ involves structural breaks, we add the trend mechanism and structural breaks as regressors, and employ OLS or FGLS depending on whether the error is uncorrelated or not.

**Spurious regression of an integrated process on a trend stationary process (I(1)-TS)**

*Case 3A (I(1)-TS regression without structural breaks).*

Consider two independent time series $y_t$ and $x_t$ generated by Case 3A in TABLE 1, where $y_t$ is an integrated process and $x_t$ is a trend stationary process. Durlauf and Phillips (1988) point out that an integrated process with a drift is spuriously correlated with a trend stationary process simply because the drift is correlated with the trend. Furthermore, an integrated process without a drift is also spuriously correlated with a trend stationary process, because the high persistency characteristics of the integrated process correlate with the trend. Noriega and Ventosa-Santaulària (2007) show the spurious regression between an integrated process and a trend stationary process (as Regression 1). To solve the problem, Noriega and Ventosa-Santaulària (2005) propose adding the linear trend as a regressor (as Regression 2), and show that the *t*-statistic $t_{\hat{\gamma}}$ possesses an asymptotic distribution which is very close to the standard normal distribution.

However, we think that Noriega and Ventosa-Santaulària (2005) neglect the possibility of autocorrelation in the OLS error. In this situation, the asymptotic distribution of the *t*-statistic can no longer be well approximated by the standard normal distribution so that the spurious relationship may still exist in Regression 2. Generally, we recommend allowing for autocorrelation in theoretical and empirical studies. Proposition 4 proves that by using FGLS, Regression 3 solves the spurious regression of an integrated process on a trend stationary process successfully.

*Proposition 4.* Suppose independent series $y_t$ and $x_t$ are generated by Case3A in TABLE 1. Running regression 3 with FGLS, we get

$$\hat{\gamma} \xrightarrow{d} N(0, \sigma_{\hat{\gamma}}^2), \quad t_{\hat{\gamma}} \xrightarrow{d} N(0,1)$$

where $\sigma_{\hat{\gamma}}^2$ depends on $x_t$.

*Proof.* See Appendix A.

Now we investigate the finite sample behavior of the *t*-statistic $t_{\hat{\gamma}}$ by Monte Carlo simulations. $y_t$ and $x_t$ are generated by Case 3A in TABLE 1, where $\varepsilon_{yt}$ and $\varepsilon_{xt}$ have standard normal distributions, $\mu_y$ is chosen to be 0.8, and the choices of other parameters are given in TABLE 7. To save some space, we do not show the results of Regression 1, which is a well-known case of spurious regression. Instead, we give the results of Regression 2, which is suggested by Noriega and Ventosa-Santaulària (2005) as a solution to spurious regression. In addition, we present the performance of our approach, Regression 3, and compare it with the results of Regression 2. The number of replications is 10,000. TABLE 7 reports the rejection rates of the null hypothesis $H_0$: $\gamma = 0$ at the 5% significant level.

TABLE 7
*Proportion of rejections of Case3A (I(1)-TS regression without structural breaks)*

| $\beta_y$ | $\beta_x$ | $\phi_x$ | Regression 2 | | | Regression 3 | | |
|---|---|---|---|---|---|---|---|---|
| | | | T=50 | T=100 | T=10000 | T=50 | T=100 | T=10000 |
| 0 | 0 | 0 | 0.0460 | 0.0481 | 0.0460 | 0.0405 | 0.0447 | 0.0447 |
| | | 0.3 | 0.1216* | 0.1294* | 0.1543* | 0.0485 | 0.0422 | 0.0548 |
| | | 0.9 | 0.4560* | 0.5453* | 0.6491* | 0.0627 | 0.0514 | 0.0490 |
| 0 | 0.2 | 0 | 0.0513 | 0.0517 | 0.0469 | 0.0415 | 0.0470 | 0.0472 |
| | | 0.3 | 0.1251* | 0.1317* | 0.1423* | 0.0432 | 0.0431 | 0.0479 |
| | | 0.9 | 0.4510* | 0.5440* | 0.6478* | 0.0607 | 0.0508 | 0.0517 |
| 0.2 | 0 | 0 | 0.0473 | 0.0470 | 0.0489 | 0.0383 | 0.0449 | 0.0530 |
| | | 0.3 | 0.1247* | 0.1358* | 0.1503* | 0.0446 | 0.0490 | 0.0543 |
| | | 0.9 | 0.4523* | 0.5495* | 0.6482* | 0.0623 | 0.0515 | 0.0522 |
| 0.2 | 0.2 | 0 | 0.0483 | 0.0478 | 0.0506 | 0.0405 | 0.0444 | 0.0537 |
| | | 0.3 | 0.1204* | 0.1315* | 0.1509* | 0.0387 | 0.0419 | 0.0522 |
| | | 0.9 | 0.4612* | 0.5493* | 0.6436* | 0.0633 | 0.0553 | 0.0505 |

Notes: * indicates that the proportion of rejections is more than 10%, which shows the phenomenon of spurious regression apparently.

TABLE 7 indicates that Regression 2 removes the spuriousness problem when an integrated process is regressed on a trend stationary process in the absence of autocorrelation. If $y_t$ does contain autocorrelation, then as $\phi$ approaches 1, the asymptotic rejection rate becomes larger. For example, the finite sample rejection rates for the case of $\beta_y=\beta_x= 0.2$ and $T=100$, are 13% and 55% for $\phi_x= 0.3$ and 0.9, respectively. On the other hand, the remaining experiments by Regression 3 reveal that the rejection rates are very close to the nominal 5% level, which suggests that the problem of spurious regression has been removed successfully.

*Case 3B (I(1)-TS regression with structural breaks).*

Consider two independent time series $y_t$ and $x_t$ generated by Case 3B in TABLE 1, where $y_t$ is an integrated process and $x_t$ is a trend stationary process. Noriega and Ventosa-Santaulària (2005) point out that when structural breaks are a feature of the data (either $y_t$ or $x_t$), the spurious relationship is present. Specifically, Regression 2 fails to solve the spurious regression when $y_t$ or $x_t$ involves structural breaks. In the same spirit of our unified approach, we tackle the problem by adding the trending mechanism and structural breaks as regressors, and using FGLS when error is autocorrelated.

Now we present the Monte Carlo performance of the *t*-statistics, using Regression 3 when $y_t$ has no structural break, and using Regression 5 when $y_t$ has structural breaks. In addition, we show the Monte Carlo results of Regression 2 for the sake of comparison. In our simulation, each of $y_t$ and $x_t$ contains at most one structural break. The trend parameters are chosen to be $\beta_x=\beta_y=0.2$. The choices of other parameters are given in TABLE 8. The break in $y_t$ is located at $T/2$, and the break in $x_t$ is at $T/5$, where $T$ is selected to be 50, 100, and 10,000. The number of replications is 10,000. TABLE 8 shows the rejection rates of the null hypothesis H$_0$: $\gamma = 0$ at the 5% significant level.

TABLE 8 shows that under Regression 2, the phenomenon of spurious regression is present,

unless $x_t$ has neither autocorrelation nor structural break. The spurious regression can be removed by applying FGLS to Regression 5 or Regression 3 depending on whether $y_t$ has structural breaks or not. To achieve even better finite sample results when $x_t$ is autocorrelated, one may as well consider using the feasible generalized median estimator.

TABLE 8
*Proportion of rejections of Case 3B (I(1)-TS regression with structural breaks)*

| $B_{1y}$ | $\beta_{1x}$ | $\phi_x$ | Regression 2 | | | Regression 3 or 5 | | |
|---|---|---|---|---|---|---|---|---|
| | | | $T=50$ | $T=100$ | $T=10000$ | $T=50$ | $T=100$ | $T=10000$ |
| 0 | 0.2 | 0 | 0.0994 | 0.3648 | 0.9591 | 0.0405 | 0.0429 | 0.0496 |
| | | 0.3 | 0.1610 | 0.3726 | 0.9642 | 0.0391 | 0.0465 | 0.0528 |
| | | 0.9 | 0.4421 | 0.5378 | 0.9605 | 0.0612 | 0.0475 | 0.0501 |
| 0.2 | 0 | 0 | 0.0507 | 0.0466 | 0.0517 | 0.0420 | 0.0427 | 0.0502 |
| | | 0.3 | 0.1283 | 0.1390 | 0.1439 | 0.0430 | 0.0444 | 0.0463 |
| | | 0.9 | 0.4575 | 0.5639 | 0.6476 | 0.0681 | 0.0547 | 0.0471 |
| 0.2 | 0.2 | 0 | 0.0876 | 0.3298 | 1.0000 | 0.0409 | 0.0406 | 0.0490 |
| | | 0.3 | 0.1542 | 0.3439 | 1.0000 | 0.0432 | 0.0446 | 0.0500 |
| | | 0.9 | 0.4533 | 0.5463 | 1.0000 | 0.0665 | 0.0527 | 0.0518 |

In summary, the spurious regression of an integrated process on a trend stationary process can be prevented by conducting an augmented FGLS regression with trending mechanism as an additional regressor. The FGLS estimator of the coefficient for the explanatory trend stationary process is proven to be asymptotically normal, and the corresponding *t*-statistic possesses an asymptotic standard normal distribution. If the dependent variable involves structural breaks, we should not only apply FGLS but also add the trend and the structural breaks as regressors in the regression.

*Remark 2.* One may argue that if the original model is not a mis-specified model, (that is $y_t = \alpha + \gamma x_t + u_t$, $\gamma \neq 0$ which reflects the true relationship between $y_t$ and $x_t$), adding the time trend as an extra regressor is likely to give rise to a spurious regression. We consider each case without structural breaks below.

For Case 1, the regression between two trend stationary variables (TS-TS), we conduct Monte Carlo simulation. The DGP of $x_t$ is a trend stationary process as the Case 1A in TABLE 1, where $\mu_x = 0.8$, $\beta_x = 0.2$, $\phi_x = 0.3$ and $\varepsilon_{xt}$ is drawn from $N(0,1)$. The original model is $y_t = 0.8 + 0.2x_t + u_{yt}$, $u_{yt}$

= $0.3u_{yt-1} + \varepsilon_{yt}$, where $\varepsilon_{yt}$ is also drawn from $N(0,1)$. The sample size and the number of replications are all 10,000. We implement FGLS regression with time as a regressor, as Regression 3. The statistics of the estimated coefficient and corresponding *t*-statistics of $x_t$ are denoted as COEF_TSTS and TSTAT_TSTS, and their histograms are graphed as Figure B1. It is clearly shown that these statistics are all normal distributions according to Jarque-Bera test. More importantly, the *t*-statistics of $x_t$ (*i.e.* TSTAT_TSTS) are large enough to reject the null hypothesis of $\gamma=0$, which means the coefficient of $x_t$ is significant. This result shows that adding trending mechanism as an extra regressor will not produce a spurious regression if it is absent in the original regression. This result is robust to the selection of the parameters of the DGPs.

For Case 2, the regression of a trend stationary process on an integrated process (TS-I(1)), the original model is necessarily mis-specified. Obviously, the combination of an integrated process $x_t$ and other stationary time series will always be an integrated process, but not a trend stationary process.

For Case 3, the regression of an integrated process on a trend stationary process (I(1)-TS), we also conduct Monte Carlo simulation. The DGP of $x_t$ is a trend stationary process as the Case 3A in TABLE 1, where $\mu_x=0.8$, $\beta_x=0.2$, $\phi_x=0.3$ and $\varepsilon_{xt}$ is drawn from $N(0,1)$. The original model is $y_t$ =$0.8+0.2x_t+u_{yt}$, $u_{yt}=u_{yt-1}+\varepsilon_{yt}$, where $\varepsilon_{yt}$ is drawn from $N(0,1)$. The sample size and the number of replications are both 10,000. We implement FGLS regression with time as a regressor, as Regression 3. The statistics of the estimated coefficient and the corresponding *t*-statistics of $x_t$ are denoted as COEF_I1TS and TSTAT_I1TS, and their histograms are graphed as Figure B2. Like the result in Case 1, the *t*-statistics of $x_t$ (*i.e.* TSTAT_I1TS) are large enough to reject the null hypothesis of $\gamma=0$, which means the coefficient of $x_t$ is significant. This result shows that the original correct model will not be identified as a spurious regression by our method. The result is also robust to the selection of the parameters of the DGPs.

## IV. An example

The time series literature on spurious regression can be traced to Yule (1926), which innovatively witnesses the spurious correlation between two independent nonstationary series graphed in Figure 1. Note that there is a significant relationship between the proportion of Church of England marriages to all marriages and the mortality rate in England and Wales over 1866-1911, as the correlation coefficient is as high as 0.9512 between the two irrelevant time series. Yule (1926) introduced the concept of spurious correlation.

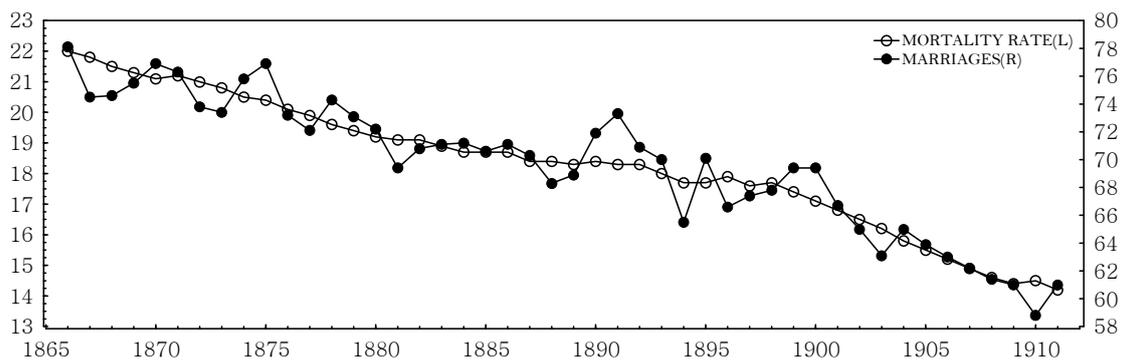

Figure 1. Mortality rate in England and Wales (left) and the proportion of Church of England marriages to all marriages (right)[1]

We now illustrate our solution to spurious regression by working on the two time series in Yule (1926). Let $y_t$ be the mortality rate in England and Wales, and $x_t$ be the proportion of Church of England marriages to all marriages. The spurious correlation between the two time series in Yule (1926) can be reflected by Regression 1. To prevent the spurious regression, García-Belmonte and Ventosa-Santaulària (2011) propose to add a deterministic trend into the regression, leading to our Regression 2. Here we suggest using FGLS (as Regression 3) to solve the spurious regression problem, considering that the autocorrelation may exist in the OLS error. The regression results of Regressions 1, 2, and 3 are presented in TABLE 9.

---

[1] The data in Figure 1 come from http://ftp.uni-bayreuth.de/math/statlib/datasets/hipl-mcleod. The correlation coefficient of the two time series in the data is 0.9515, slightly different from the correlation coefficient given by Yule, G. U. (1926). 'Why do we sometimes get nonsense-correlations between Time-Series?--a study in sampling and the nature of time-series', *Journal of the royal statistical society,* Vol. 89, pp. 1-63.. All the percentage values are times 1000 for clarity.

TABLE 9
*Regression coefficients and statistics of Regression 1, 2, and 3*

|  | Regression 1 | Regression 2 | Regression 3 |
|---|---|---|---|
| $\hat{\alpha}$ | -10.8466‡ | 12.4418‡ | 22.5030‡ |
|  | (-7.5145) | (5.8182) | (8.4931) |
| $\hat{\gamma}$ | 0.4185‡ | 0.1217‡ | 0.0018 |
|  | (20.5251) | (4.3999) | (0.1468) |
| $\hat{\beta}$ | --- | -0.1155‡ | -0.1823‡ |
|  |  | (-11.5484) | (-3.5094) |
| $\hat{\rho}$ | --- | --- | 0.9526‡ |
|  |  |  | (15.7573) |
| $Adj-R^2$ | 0.9033 | 0.9759 | 0.9946 |
| $DW$ | 1.5356 | 0.7158 | 1.7786 |
| $Q_{(10)}$ | 12.598 | 47.1340 | 8.6880 |

Notes: --- denotes the nonexistence of the coefficient in the corresponding regression. The numbers in the parentheses denote the t-statistics. ‡ denotes the statistical significance of the coefficient at the 1% significant level.

From TABLE 9, it can be seen that the estimate for $\gamma$ under Regression 1 is 0.4185. Its associated *t*-statistic, 20.5251, has statistical significance at the 1% level. Thus, Regression 1 claims that the mortality rate in England and Wales can be explained by the proportion of Church of England marriages and thus the spurious relationship is present. Regression 2, proposed by García-Belmonte and Ventosa-Santaulària (2011), gives an estimate 0.1217 for $\gamma$ and its *t*-statistic 4.399. Although this *t*-statistic from Regression 2 is much smaller than that based on Regression 1, it is still statistically significant at the 1% level, and so the spurious relationship is still present in Regression 2. It is also noteworthy that *Durbin-Watson* statistic based on Regression 2 is 0.7158, which indicates strong positive autocorrelation. To deal with this autocorrelation, we apply Regression 3. Regression 3 not only handles the impact of trends by adding the trending mechanism as a regressor, but also solves the problem of autocorrelation by using FGLS. From Regression 3, the estimator of $\gamma$ is 0.0018, and its corresponding *t*-statistic is 0.1468, which means the phenomenon of spurious regression disappears. Hence, Regression 3 provides evidence on the irrelevance of the proportion of Church of England marriages to the mortality rate in England and Wales. In addition, Regression 3 has an adjusted $R^2$ value (0.9946) that is higher than that based on Regression 1 (0.9033) and Regression 2 (0.9759), and the *Durbin-Watson* statistic is 1.7786, which indicates that the autocorrelation problem in Regression 2 has been corrected. To sum up, we can address the problem of spurious regression successfully by running Regression 3.

## V. Conclusion

Spurious regression can arise wherever there is a trending mechanism in the DGP. This paper considers three types of spurious regressions caused by trending mechanisms, i.e., regressions between two trend stationary processes, regressions of a trend stationary process on an integrated process, and regressions of an integrated process on a trend stationary process. Each type is divided into two cases depending on whether structural breaks are allowed or not. We show that spurious regression is present in all cases.

In our view, the problem of spurious regression with trending variables is caused by the omission of important explanatory variables (as trending mechanism and structural breaks) and the existence of autocorrelation. When we doubt that some regression between two trending variables is a spurious regression, by adding trending mechanism as an explanatory variable, applying FGLS or OLS depending on whether autocorrelation exists or not, and including structural break variables in the regression when breaks occur, spurious regression problem can be removed successfully. In this paper, we focus on tackling the spurious regression problem, so we assume the information of structural breaks are known a prior. In practical, we could determine the number, the types, and the locations of the structural breaks by existing methods in advance.

Our approach has been justified theoretically, as the *t*-statistic testing the correlation between $y_t$ and $x_t$ converges weakly to either the standard normal distribution or a distribution that is very close to the standard normal distribution. In addition, the validity of the approach has been confirmed by Monte Carlo simulations. Moreover, the spurious connection between the two time series in Yule (1926) has been successfully removed by applying our methods, that is, by adding trending mechanism as an explanatory variable and using FGLS.

## Appendix A: Proofs

**Proof of Proposition 1:** We apply the proposition and corollary in the section 3 of Rothenberg (1984). Let the vector $c = (0,0,1)'$, then $\hat{\gamma} \xrightarrow{d} N(0, \sigma_{\hat{\gamma}}^2)$, where $\sigma_{\hat{\gamma}}^2$ depends on $\phi_y$ and $x_t$, and then $t_{\hat{\gamma}} \xrightarrow{d} N(0,1)$.

**Proof of Proposition 2:** Let $x_{2t} = t$, $x_{3t} = x_t$. Define the demeaned time series as follows:

$$\dot{x}_{2t} = t - \frac{T+1}{2}, \dot{x}_{3t} = x_{3t} - \frac{1}{T}\sum x_{3t},$$

$$\dot{y}_t = (\alpha + \beta t + \varepsilon_{yt}) - \frac{1}{T}\sum(\alpha + \beta t + \varepsilon_{yt}) = \beta(t - \frac{T+1}{2}) + \varepsilon_{yt} - \frac{1}{T}\sum \varepsilon_{yt}.$$

Notice that $\hat{\gamma}_1 = \dfrac{(\sum \dot{y}_t \dot{x}_{3t})(\sum \dot{x}_{2t}^2) - (\sum \dot{y}_t \dot{x}_{2t})(\sum \dot{x}_{2t}\dot{x}_{3t})}{(\sum \dot{x}_{2t}^2)(\sum \dot{x}_{3t}^2) - (\sum \dot{x}_{2t}\dot{x}_{3t})^2}$, with

$$\sum \dot{x}_{2t}\dot{x}_{3t} = \sum t\dot{x}_{3t} - \frac{T+1}{2}\sum \dot{x}_{3t} = \sum tx_{3t} - \frac{T+1}{2}\sum x_{3t},$$

$$\sum \dot{x}_{2t}^2 = \sum (t - \frac{T+1}{2})^2 = \frac{(T-1)T(T+1)}{12},$$

$$\sum \dot{y}_t \dot{x}_{3t} = \beta(\sum tx_{3t} - \frac{T+1}{2}\sum x_{3t}) + \sum \varepsilon_{yt} x_{3t} - \frac{1}{T}(\sum_{1}^{T} \varepsilon_{yt} \sum x_{3t}),$$

$$\sum \dot{y}_t \dot{x}_{2t} = \beta \frac{(T-1)T(T+1)}{12} + \sum t\varepsilon_{yt} - \frac{T+1}{2}\sum \varepsilon_{yt},$$

$$\sum \dot{x}_{3t}^2 = \sum(x_{3t} - \frac{1}{T}\sum x_{3t})^2 = \sum x_{3t}^2 - \frac{1}{T}(\sum x_{3t})^2.$$

Then $T^{-5}[(\sum \dot{x}_{2t}^2)(\sum \dot{x}_{3t}^2) - (\sum \dot{x}_{2t}\dot{x}_{3t})^2] \Rightarrow \frac{1}{12}\sigma_w^2\left\{\int_0^1 W(t)^2 dt - [\int_0^1 W(t)dt]^2\right\} - \sigma_w^2\left\{\int_0^1 tW(t)dt - \frac{1}{2}\int_0^1 W(t)dt\right\}^2$.

On the other hand,

$$(\sum \dot{y}_t \dot{x}_{3t})(\sum \dot{x}_{2t}^2) - (\sum \dot{y}_t \dot{x}_{2t})(\sum \dot{x}_{2t}\dot{x}_{3t})$$
$$= [\frac{(T-1)T(T+1)}{12}][\sum \varepsilon_{yt} x_{3t} - \frac{1}{T}(\sum \varepsilon_{yt} \sum x_{3t})] - [\sum t\varepsilon_{yt} - \frac{T+1}{2}\sum \varepsilon_{yt}][\sum tx_{3t} - \frac{T+1}{2}\sum x_{3t}],$$

$$T^{-4}[(\sum \dot{y}_t \dot{x}_{3t})(\sum \dot{x}_{2t}^2) - (\sum \dot{y}_t \dot{x}_{2t})(\sum \dot{x}_{2t}\dot{x}_{3t})]$$
$$\Rightarrow \frac{1}{12}\sigma_v \sigma_w [\int_0^1 W(t)dV(t) - V(1)\int_0^1 W(t)dt] - \sigma_w \sigma_v[\frac{1}{2}V(1) - \int_0^1 V(t)dt][\int_0^1 tW(t)dt - \frac{1}{2}\int_0^1 W(t)dt].$$

Hence,

$$T\hat{\gamma} \Rightarrow \frac{\frac{1}{12}\sigma_v[\int_0^1 W(t)dV(t) - V(1)\int_0^1 W(t)dt] - \sigma_w \sigma_v[\frac{1}{2}V(1) - \int_0^1 V(t)dt][\int_0^1 tW(t)dt - \frac{1}{2}\int_0^1 W(t)dt]}{\frac{1}{12}\sigma_w\left\{\int_0^1 W(t)^2 dt - [\int_0^1 W(t)dt]^2\right\} - \sigma_w\left\{\int_0^1 tW(t)dt - \frac{1}{2}\int_0^1 W(t)dt\right\}^2}.$$

So

$$t_{\hat{\gamma}} = \frac{\hat{\gamma}}{se(\hat{\gamma})} = \frac{\frac{(\sum \dot{y}_t \dot{x}_{3t})(\sum \dot{x}_{2t}^2) - (\sum \dot{y}_t \dot{x}_{2t})(\sum \dot{x}_{2t} \dot{x}_{3t})}{(\sum \dot{x}_{2t}^2)(\sum \dot{x}_{3t}^2) - (\sum \dot{x}_{2t} \dot{x}_{3t})^2}}{\sigma_v \sqrt{\sum \dot{x}_{2t}^2 / [(\sum \dot{x}_{2t}^2)(\sum \dot{x}_{3t}^2) - (\sum \dot{x}_{2t} \dot{x}_{3t})^2]}} = \frac{(\sum \dot{y}_t \dot{x}_{3t})(\sum \dot{x}_{2t}^2) - (\sum \dot{y}_t \dot{x}_{2t})(\sum \dot{x}_{2t} \dot{x}_{3t})}{\sigma_v \sqrt{\sum \dot{x}_{2t}^2 [(\sum \dot{x}_{2t}^2)(\sum \dot{x}_{3t}^2) - (\sum \dot{x}_{2t} \dot{x}_{3t})^2]}}$$

$$\Rightarrow \frac{\frac{1}{12}[\int_0^1 W(t)dV(t) - V(1)\int_0^1 W(t)dt] - [\frac{1}{2}V(1) - \int_0^1 V(t)dt][\int_0^1 tW(t)dt - \frac{1}{2}\int_0^1 W(t)dt]}{\sqrt{\frac{1}{12^2}\{\int_0^1 W(t)^2 dt - [\int_0^1 W(t)dt]^2\} - \frac{1}{12}\{\int_0^1 tW(t)dt - \frac{1}{2}\int_0^1 W(t)dt\}^2}}.$$

The above $\Rightarrow$ denotes convergence in distribution, $W(t)$ and $V(t)$ are independent Wiener Processes on $C[0,1]$.

**Proof of Proposition 3:** Let $\hat{\rho}$ is the estimator of $\phi_y$, and $x_{2t} = t$, $x_{3t} = x_t$. Then

$$\tilde{x}_{2t} = t - \hat{\rho}(t-1) = \hat{\rho} + (1-\hat{\rho})t, \quad \tilde{x}_{3t} = x_t - \hat{\rho}x_{t-1} = \beta_x + x_{t-1} + \varepsilon_{xt} - \hat{\rho}x_{t-1} = \beta_x + (1-\hat{\rho})x_{t-1} + \varepsilon_{xt},$$

$$\tilde{y}_t = y_t - \hat{\rho}y_{t-1} = \mu_y(1-\hat{\rho}) + \beta_y\hat{\rho} + \beta_y(1-\hat{\rho})t + (\phi_y - \hat{\rho})\mu_{yt-1} + \varepsilon_{yt}.$$

The demeaned time series have

$$\dot{\tilde{x}}_{2t} = (1-\hat{\rho})(t - \frac{T+2}{2}),$$

$$\dot{\tilde{x}}_{3t} = (1-\hat{\rho})(x_{t-1} - \frac{1}{T-1}\sum x_{t-1}) + (\varepsilon_{xt} - \frac{1}{T-1}\sum \varepsilon_{xt})$$
$$= (1-\hat{\rho})\beta_x(t - \frac{T+2}{2}) + (1-\hat{\rho})(W_{t-1} - \frac{1}{T-1}\sum W_{t-1}) + (\varepsilon_{xt} - \frac{1}{T-1}\sum \varepsilon_{xt}),$$

where $W_t = \sum \varepsilon_{xt}$, and $\dot{\tilde{y}}_t = \beta_y(1-\hat{\rho})(t - \frac{T+2}{2}) + (\phi_y - \hat{\rho})(\mu_{yt-1} - \frac{1}{T-1}\sum \mu_{yt-1}) + (\varepsilon_{yt} - \frac{1}{T-1}\sum \varepsilon_{yt}).$

Notice that,

$$\hat{\gamma} = \frac{(\sum \dot{\tilde{y}}_t \dot{\tilde{x}}_{3t})(\sum \dot{\tilde{x}}_{2t}^2) - (\sum \dot{\tilde{y}}_t \dot{\tilde{x}}_{2t})(\sum \dot{\tilde{x}}_{2t} \dot{\tilde{x}}_{3t})}{[(\sum \dot{\tilde{x}}_{2t}^2)(\sum \dot{\tilde{x}}_{3t}^2) - (\sum \dot{\tilde{x}}_{2t} \dot{\tilde{x}}_{3t})^2]},$$

$$\sum \dot{\tilde{x}}_{2t}^2 = (1-\hat{\rho})^2 \frac{(T-1)T(T-2)}{12},$$

$$\sum \dot{\tilde{x}}_{3t}^2 = (1-\hat{\rho})^2 \sum (x_{t-1} - \frac{1}{T-1}\sum x_{t-1})^2 + \sum (\varepsilon_{xt} - \frac{1}{T-1}\sum \varepsilon_{xt})^2 + 2(1-\hat{\rho})\sum (x_{t-1} - \frac{1}{T-1}\sum x_{t-1})(\varepsilon_{xt} - \frac{1}{T-1}\sum \varepsilon_{xt}),$$

$$\sum \dot{\tilde{x}}_{2t}\dot{\tilde{x}}_{3t} = (1-\hat{\rho})^2 \beta_x \frac{(T-1)T(T-2)}{12} + (1-\hat{\rho})^2[\sum tW_{t-1} - \frac{T+2}{2}\sum W_{t-1}] + (1-\hat{\rho})[\sum t\varepsilon_{xt} - \frac{T+2}{2}\sum \varepsilon_{xt}],$$

$$\sum \dot{\tilde{y}}_t\dot{\tilde{x}}_{2t} = \beta_y(1-\hat{\rho})^2 \frac{(T-1)T(T-2)}{12} + (1-\hat{\rho})(\phi_y - \hat{\rho})[\sum t\mu_{yt-1} - \frac{T+2}{2}\sum \mu_{yt-1}] + (1-\hat{\rho})[\sum t\varepsilon_{yt} - \frac{T+2}{2}\sum \varepsilon_{yt}],$$

$$\sum \dot{\tilde{y}}_t\dot{\tilde{x}}_{3t} = (1-\hat{\rho})^2 \beta_x\beta_y \frac{(T-1)T(T-2)}{12} + (1-\hat{\rho})(\phi_y - \hat{\rho})\beta_x \sum (t\mu_{yt-1} - \frac{T+2}{2}\sum \mu_{yt-1}) + (1-\hat{\rho})\beta_x \sum (t\varepsilon_{yt} - \frac{T+2}{2}\sum \varepsilon_{yt})$$
$$+ (1-\hat{\rho})^2 \beta_y \sum (tW_{t-1} - \frac{T+2}{2}\sum W_{t-1}) + (1-\hat{\rho})(\phi_y - \hat{\rho})\sum (\mu_{yt-1} - \frac{1}{T-1}\sum \mu_{yt-1})(W_{t-1} - \frac{1}{T-1}\sum W_{t-1})$$
$$+ (1-\hat{\rho})\sum (\varepsilon_{yt} - \frac{1}{T-1}\sum \varepsilon_{yt})(W_{t-1} - \frac{1}{T-1}\sum W_{t-1}) + (1-\hat{\rho})\beta_y \sum (t\varepsilon_{xt} - \frac{T+2}{2}\sum \varepsilon_{xt})$$
$$+ (\phi_y - \hat{\rho})\sum (\mu_{yt-1} - \frac{1}{T-1}\sum \mu_{yt-1})(\varepsilon_{xt} - \frac{1}{T-1}\sum \varepsilon_{xt}) + \sum (\varepsilon_{yt} - \frac{1}{T-1}\sum \varepsilon_{yt})(\varepsilon_{xt} - \frac{1}{T-1}\sum \varepsilon_{xt})$$

Then,

$$(\sum \dot{\tilde{x}}_{2t}^2)(\sum \dot{\tilde{x}}_{3t}^2) - (\sum \dot{\tilde{x}}_{2t}\dot{\tilde{x}}_{3t})^2$$
$$= (1-\hat{\rho})^2 \frac{(T-1)T(T-2)}{12} \times \{(1-\hat{\rho})^2 \sum (x_{t-1} - \frac{1}{T-1}\sum x_{t-1})^2 + \sum (\varepsilon_{xt} - \frac{1}{T-1}\sum \varepsilon_{xt})^2$$
$$+ 2(1-\hat{\rho})\sum (x_{t-1} - \frac{1}{T-1}\sum x_{t-1})(\varepsilon_{xt} - \frac{1}{T-1}\sum \varepsilon_{xt})\} - \{(1-\hat{\rho})^2[\sum tx_{t-1} - \frac{T+2}{2}\sum x_{t-1}] + (1-\hat{\rho})[\sum t\varepsilon_{xt} - \frac{T+2}{2}\sum \varepsilon_{xt}]\}^2$$

$$(\sum \dot{\tilde{y}}_t \dot{\tilde{x}}_{3t})(\sum \dot{\tilde{x}}_{2t}^2) - (\sum \dot{\tilde{y}}_t \dot{\tilde{x}}_{2t})(\sum \dot{\tilde{x}}_{2t} \dot{\tilde{x}}_{3t})$$

$$= [(1-\hat{\rho})^2 \frac{(T-1)T(T-2)}{12}] \times \{(1-\hat{\rho})(\phi_y - \hat{\rho})\sum(\mu_{yt-1} - \frac{1}{T-1}\sum \mu_{yt-1})(W_{t-1} - \frac{1}{T-1}\sum W_{t-1})$$

$$+(1-\hat{\rho})\sum(\varepsilon_{yt} - \frac{1}{T-1}\sum \varepsilon_{yt})(W_{t-1} - \frac{1}{T-1}\sum W_{t-1}) + (\phi_y - \hat{\rho})\sum(\mu_{yt-1} - \frac{1}{T-1}\sum \mu_{yt-1})(\varepsilon_{xt} - \frac{1}{T-1}\sum \varepsilon_{xt})\}$$

$$+\sum(\varepsilon_{yt} - \frac{1}{T-1}\sum \varepsilon_{yt})(\varepsilon_{xt} - \frac{1}{T-1}\sum \varepsilon_{xt})\} - \{(1-\hat{\rho})(\phi_y - \hat{\rho})[\sum t\mu_{yt-1} - \frac{T+2}{2}\sum \mu_{yt-1}] + (1-\hat{\rho})[\sum t\varepsilon_{yt} - \frac{T+2}{2}\sum \varepsilon_{yt}]\}$$

$$\times\{(1-\hat{\rho})^2[\sum tW_{t-1} - \frac{T+2}{2}\sum W_{t-1}] + (1-\hat{\rho})[\sum t\varepsilon_{xt} - \frac{T+2}{2}\sum \varepsilon_{xt}]\}$$

By Grenander and Rosenblatt (1957), $T^{1/2}(\hat{\rho} - \phi_y) \xrightarrow{d} N(0, 1-\phi_y^2)$,

$$T^{-5}(\sum \dot{\tilde{x}}_{2t}^2)(\sum \dot{\tilde{x}}_{3t}^2) - (\sum \dot{\tilde{x}}_{2t} \dot{\tilde{x}}_{3t})^2 \Rightarrow \sigma_w^2(1-\phi_y)^4 \frac{1}{12}\{[\int_0^1 W(t)^2 dt - [\int_0^1 W(t)dt]^2] - [\int_0^1 tW(t)dt - \frac{1}{2}\int_0^1 W(t)dt]^2\}.$$

$$T^{-4}(\sum \dot{\tilde{y}}_t \dot{\tilde{x}}_{3t})(\sum \dot{\tilde{x}}_{2t}^2) - (\sum \dot{\tilde{y}}_t \dot{\tilde{x}}_{2t})(\sum \dot{\tilde{x}}_{2t} \dot{\tilde{x}}_{3t})$$

$$\Rightarrow (1-\phi_y)^3 \sigma_y \sigma_w \frac{1}{12}\{[\int_0^1 W(t)dV(t) - V(1)\int_0^1 W(t)dt] - [\frac{1}{2}V(1) - \int_0^1 V(t)dt][\int_0^1 tW(t)dt - \frac{1}{2}\int_0^1 W(t)dt]\}.$$

So,

$$T\hat{\gamma} \Rightarrow \frac{\sigma_y \{\frac{1}{12}[\int_0^1 W(t)dV(t) - V(1)\int_0^1 W(t)dt] - [\frac{1}{2}V(1) - \int_0^1 V(t)dt][\int_0^1 tW(t)dt - \frac{1}{2}\int_0^1 W(t)dt]\}}{((1-\phi_y)\sigma_w)\{\frac{1}{12}\{\int_0^1 W(t)^2 dt - [\int_0^1 W(t)dt]^2\} - \{\int_0^1 tW(t)dt - \frac{1}{2}\int_0^1 W(t)dt\}^2\}},$$

$$t_{\hat{\gamma}} = \frac{\hat{\gamma}}{se(\hat{\gamma})} = \frac{\frac{(\sum \dot{\tilde{y}}_t \dot{\tilde{x}}_{3t})(\sum \dot{\tilde{x}}_{2t}^2) - (\sum \dot{\tilde{y}}_t \dot{\tilde{x}}_{2t})(\sum \dot{\tilde{x}}_{2t} \dot{\tilde{x}}_{3t})}{[(\sum \dot{\tilde{x}}_{2t}^2)(\sum \dot{\tilde{x}}_{3t}^2) - (\sum \dot{\tilde{x}}_{2t} \dot{\tilde{x}}_{3t})^2]}}{\sigma_y \sqrt{\sum \dot{\tilde{x}}_{2t}^2 / [(\sum \dot{\tilde{x}}_{2t}^2)(\sum \dot{\tilde{x}}_{3t}^2) - (\sum \dot{\tilde{x}}_{2t} \dot{\tilde{x}}_{3t})^2]}} = \frac{(\sum \dot{\tilde{y}}_t \dot{\tilde{x}}_{3t})(\sum \dot{\tilde{x}}_{2t}^2) - (\sum \dot{\tilde{y}}_t \dot{\tilde{x}}_{2t})(\sum \dot{\tilde{x}}_{2t} \dot{\tilde{x}}_{3t})}{\sigma_y \sqrt{\sum \dot{\tilde{x}}_{2t}^2[(\sum \dot{\tilde{x}}_{2t}^2)(\sum \dot{\tilde{x}}_{3t}^2) - (\sum \dot{\tilde{x}}_{2t} \dot{\tilde{x}}_{3t})^2]}}$$

$$\Rightarrow \frac{\{\frac{1}{12}[\int_0^1 W(t)dV(t) - V(1)\int_0^1 W(t)dt] - [\frac{1}{2}V(1) - \int_0^1 V(t)dt][\int_0^1 tW(t)dt - \frac{1}{2}\int_0^1 W(t)dt]\}}{\sqrt{\frac{1}{12^2}\{\int_0^1 W(t)^2 dt - [\int_0^1 W(t)dt]^2\} - \frac{1}{12}\{\int_0^1 tW(t)dt - \frac{1}{2}\int_0^1 W(t)dt\}^2}}$$

The above $\Rightarrow$ denotes convergence in distribution, $W(t)$ and $V(t)$ are independent Wiener Processes on $C[0,1]$.

**Proof of Proposition 4:** By Phillips and Hodgson (1994) and Giraitis and Phillips (2012), $T(1-\hat{\rho}) = O_p(1)$.

Let $x_{2t} = t, x_{3t} = x_t$, $v_t = u_{xt}$, and let

$$\tilde{y}_t = y_t - \hat{\rho}y_{t-1} = \beta_y + (1-\hat{\rho})y_{t-1} + \varepsilon_{yt},$$

$$\tilde{x}_{2t} = \hat{\rho} + (1-\hat{\rho})t, \quad \tilde{x}_{3t} = (1-\hat{\rho})\mu_x + \beta_x \hat{\rho} + (1-\hat{\rho})\beta_x t + (v_t - \hat{\rho}v_{t-1}).$$

The demeaned time series are given by

$$\dot{\tilde{y}}_t = (1-\hat{\rho})(y_{t-1} - \frac{1}{T-1}\sum y_{t-1}) + (\varepsilon_{yt} - \frac{1}{T-1}\sum \varepsilon_{yt}), \quad \dot{\tilde{x}}_{2t} = (1-\hat{\rho})(t - \frac{T+2}{2}),$$

$$\dot{\tilde{x}}_{3t} = (1-\hat{\rho})\beta_x(t - \frac{T+2}{2}) + (v_t - \frac{1}{T-1}\sum v_t) - \hat{\rho}(v_{t-1} - \frac{1}{T-1}\sum v_{t-1}).$$

Notice that,

$$\sum \dot{\tilde{x}}_{2t}^2 = (1-\hat{\rho})^2 \frac{(T-1)T(T-2)}{12},$$

$$\sum \dot{\tilde{x}}_{3t}^2 = (1-\hat{\rho})^2 \beta_x^2 \frac{(T-1)T(T-2)}{12} + \sum v_t^2 - \frac{1}{T-1}(\sum v_t)^2 + \hat{\rho}^2(\sum v_{t-1}^2 - \frac{1}{T-1}(\sum v_{t-1})^2)$$

$$+ 2(1-\hat{\rho})\beta_x(\sum tv_t - \frac{T+2}{2}\sum v_t) - 2(1-\hat{\rho})\hat{\rho}\beta_x(\sum tv_{t-1} - \frac{T+2}{2}\sum v_{t-1}) - 2\hat{\rho}(\sum v_t v_{t-1} - \frac{1}{T-1}\sum v_t \sum v_{t-1}),$$

$$\sum \dot{\tilde{x}}_{2t}\dot{\tilde{x}}_{3t} = (1-\hat{\rho})^2 \beta_x \frac{(T-1)T(T-2)}{12} + (1-\hat{\rho})(\sum tv_t - \frac{T+2}{2}\sum v_t) - (1-\hat{\rho})\hat{\rho}(\sum tv_{t-1} - \frac{T+2}{2}\sum v_{t-1}),$$

$$\sum \dot{y}_t \dot{\tilde{x}}_{2t} = (1-\hat{\rho})^2 (\sum t y_{t-1} - \frac{T+2}{2}\sum y_{t-1}) + (1-\hat{\rho})(\sum t\varepsilon_{yt} - \frac{T+2}{2}\sum \varepsilon_{yt}),$$

$$\sum \dot{y}_t \dot{\tilde{x}}_{3t} = (1-\hat{\rho})^2 \beta_x (\sum t y_{t-1} - \frac{T+2}{2}\sum y_{t-1}) + (1-\hat{\rho})\beta_x (\sum t\varepsilon_{yt} - \frac{T+2}{2}\sum \varepsilon_{yt})$$
$$+ (1-\hat{\rho})\sum(y_{t-1} - \frac{1}{T-1}\sum y_{t-1})(v_t - \frac{1}{T-1}\sum v_t) + \sum(v_t - \frac{1}{T-1}\sum v_t)(\varepsilon_{yt} - \frac{1}{T-1}\sum \varepsilon_{yt}))$$
$$- (1-\hat{\rho})\hat{\rho}\sum(y_{t-1} - \frac{1}{T-1}\sum y_{t-1})(v_{t-1} - \frac{1}{T-1}\sum v_{t-1}) + \hat{\rho}\sum(v_{t-1} - \frac{1}{T-1}\sum v_{t-1})(\varepsilon_{yt} - \frac{1}{T-1}\sum \varepsilon_{yt}))$$

Let $(1-\hat{\rho}) = \frac{c}{T}$, $c$ is a constant, then

$$(\sum \dot{\tilde{x}}_{2t}^2)(\sum \dot{\tilde{x}}_{3t}^2) - (\sum \dot{\tilde{x}}_{2t}\dot{\tilde{x}}_{3t})^2$$
$$= \frac{c^2(T-1)(T-2)}{12T} \times [\sum(\Delta v_t - \frac{1}{T-1}\sum \Delta v_t)^2 + \frac{c^2}{T^2}\sum v_{t-1}^2 + \frac{2c}{T}\sum \Delta v_t v_{t-1} - \frac{1}{T-1}\frac{c^2}{T^2}(\sum v_{t-1})^2 - \frac{1}{T-1}\frac{2c}{T}\sum \Delta v_t \sum v_{t-1}]$$
$$- \frac{c^2}{T^2}[(\sum t\Delta v_t - \frac{T+2}{2}\sum \Delta v_t)^2 + \frac{c^2}{T}(\sum tv_{t-1} - \frac{T+2}{2}\sum v_{t-1})^2 + 2c(\sum t\Delta v_t - \frac{T+2}{2}\sum \Delta v_t)(\sum tv_{t-1} - \frac{T+2}{2}\sum v_{t-1})]$$

$$(\sum \dot{y}_t \dot{\tilde{x}}_{3t})(\sum \dot{\tilde{x}}_{2t}^2) - (\sum \dot{y}_t \dot{\tilde{x}}_{2t})(\sum \dot{\tilde{x}}_{2t}\dot{\tilde{x}}_{3t})$$
$$= [(1-\hat{\rho})^2 \frac{(T-1)T(T-2)}{12}] \times \{(1-\hat{\rho})\sum(y_{t-1} - \frac{1}{T-1}\sum y_{t-1})(\Delta v_t - \frac{1}{T-1}\sum \Delta v_t) + \sum(\Delta v_t - \frac{1}{T-1}\sum \Delta v_t)(\varepsilon_{yt} - \frac{1}{T-1}\sum \varepsilon_{yt})$$
$$- (1-\hat{\rho})(\hat{\rho}-1)\sum(y_{t-1} - \frac{1}{T-1}\sum y_{t-1})(v_{t-1} - \frac{1}{T-1}\sum v_{t-1}) - (\hat{\rho}-1)\sum(v_{t-1} - \frac{1}{T-1}\sum v_{t-1})(\varepsilon_{yt} - \frac{1}{T-1}\sum \varepsilon_{yt})\}$$
$$- [(1-\hat{\rho})^2(\sum ty_{t-1} - \frac{T+2}{2}\sum y_{t-1}) + (1-\hat{\rho})(\sum t\varepsilon_{yt} - \frac{T+2}{2}\sum \varepsilon_{yt})]$$
$$\times \{(1-\hat{\rho})(\sum tv_t - \frac{T+2}{2}\sum v_t) - (1-\hat{\rho})\hat{\rho}(\sum tv_{t-1} - \frac{T+2}{2}\sum v_{t-1})\}$$

And

$$\frac{1}{T}[(\sum \dot{\tilde{x}}_{2t}^2)(\sum \dot{\tilde{x}}_{3t}^2) - (\sum \dot{\tilde{x}}_{2t}\dot{\tilde{x}}_{3t})^2] \to \frac{c^2}{12}[\sum(\Delta v_t - \overline{(\Delta v_t)})^2] + O_p(1) + o_p(1),$$

$$\frac{1}{T}[(\sum \dot{y}_t \dot{\tilde{x}}_{3t})(\sum \dot{\tilde{x}}_{2t}^2) - (\sum \dot{y}_t \dot{\tilde{x}}_{2t})(\sum \dot{\tilde{x}}_{2t}\dot{\tilde{x}}_{3t})] \to \frac{c^2}{12}[\sum(\varepsilon_{yt} - \overline{\varepsilon_{yt}})(\Delta v_t - \overline{(\Delta v_t)})] + +O_p(1) + o_p(1).$$

Hence, $\hat{\gamma} = \frac{(\sum \dot{y}_t \dot{\tilde{x}}_{3t})(\sum \dot{\tilde{x}}_{2t}^2) - (\sum \dot{y}_t \dot{\tilde{x}}_{2t})(\sum \dot{\tilde{x}}_{2t}\dot{\tilde{x}}_{3t})}{[(\sum \dot{\tilde{x}}_{2t}^2)(\sum \dot{\tilde{x}}_{3t}^2) - (\sum \dot{\tilde{x}}_{2t}\dot{\tilde{x}}_{3t})^2]} \xrightarrow{d} N(0, \sigma_\gamma^2)$.

Further, $t_{\hat{\gamma}} \xrightarrow{d} N(0,1)$.

## Appendix B: Supplementary results

TABLE B1
*Comparison of the rejection rates (TS-TS regression without structural breaks)*

| $\beta_y$ | $\beta_x$ | $\phi_y$ | $\phi_x$ | Regression 2 | Regression 2(NW) | Regression 3 |
|---|---|---|---|---|---|---|
| 0.2 | 0 | 0 | 0 | 0.0518 | 0.0800 | 0.0541 |
| | | 0.3 | 0.3 | 0.0769 | 0.0972 | 0.0579 |
| | | 0.9 | 0.9 | 0.4182* | 0.3171* | 0.0669 |
| | | 0 | 0.9 | 0.0474 | 0.0966 | 0.0625 |
| | | 0.9 | 0 | 0.0503 | 0.0851 | 0.0434 |
| 0.2 | 0.2 | 0 | 0 | 0.0524 | 0.0761 | 0.0542 |
| | | 0.3 | 0.3 | 0.0752 | 0.0990 | 0.0548 |
| | | 0.9 | 0.9 | 0.4095* | 0.3116* | 0.0662 |
| | | 0 | 0.9 | 0.0486 | 0.0999 | 0.0677 |
| | | 0.9 | 0 | 0.0522 | 0.0837 | 0.0405 |

Notes: * indicates the proportion of rejection which is greater than 10% and illustrate the phenomenon of spurious regression.

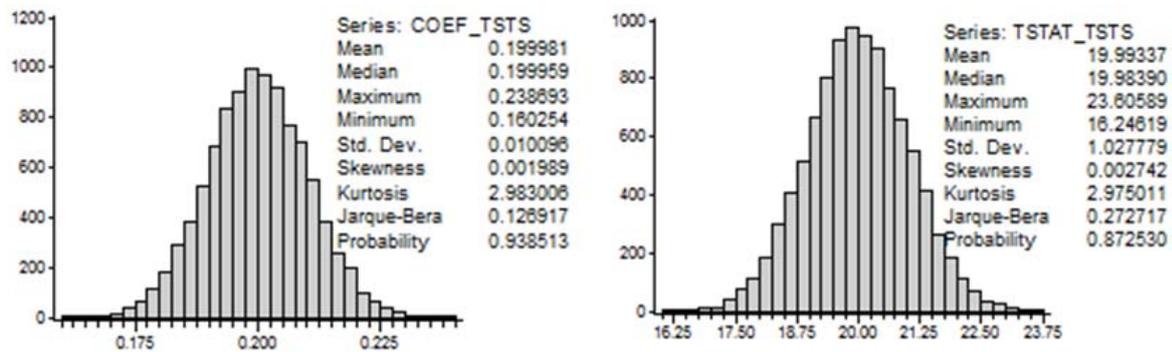
**Figure B1.** The estimated coefficient and corresponding t-statistics of $x_t$ under the TS-TS regression

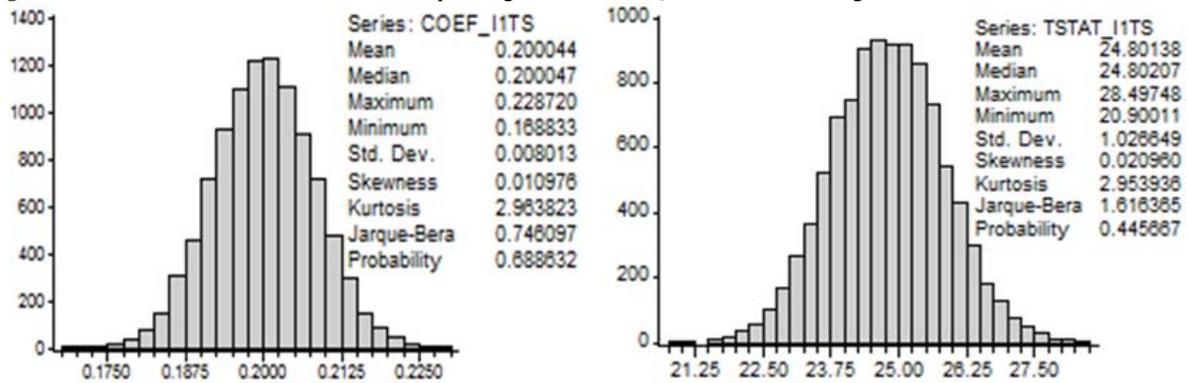
**Figure B2.** The estimated coefficient and corresponding t-statistics of $x_t$ under the I(1)-TS regression